\documentclass[a4paper,fleqn,usenatbib]{mnras}

\usepackage{newtxtext,newtxmath}
\usepackage[T1]{fontenc}
\usepackage{ae,aecompl}
\usepackage{float}

\usepackage{graphicx}	% Including figure files
\usepackage{amsmath}	% Advanced maths commands
\usepackage{amssymb}	% Extra maths symbols

\newcommand{\as}{$^{\prime\prime}$}
\newcommand\laurel[1]{EPIC~205718330}
\newcommand\hardey[1]{EPIC~235240266}

\title[Little Dippers]{The Little Dippers: Transits of Star-grazing Exocomets?}

\author[M. Ansdell et al.]{M. Ansdell,$^{1,2}$\thanks{E-mail: ansdell@berkeley.edu} E. Gaidos,$^{3}$ T. L. Jacobs,$^{4}$ A. Mann,$^{5,6}$ C. F. Manara,$^{7}$  \newauthor G. M. Kennedy,$^{8, 9}$ A. Vanderburg,$^{10,11}$ M. Kenworthy,$^{12}$ T. Hirano,$^{13,14}$ \newauthor D. M. LaCourse$^{15}$, C. Hedges$^{16}$, A. Frasca$^{17}$ \\
$^{1}$Center for Integrative Planetary Science, University of California at Berkeley, Berkeley, CA 94720, USA\\
$^{2}$Department of Astronomy, University of California at Berkeley, Berkeley, CA 94720, USA\\
$^{3}$Department of Geology \& Geophysics, University of Hawai`i at M\={a}noa, Honolulu, HI, USA\\
$^{4}$Amateur Astronomer, 12812 SE 69th Place Bellevue, WA 98006\\
$^{5}$Columbia University, Department of Astronomy, 550 West 120th Street, New York, NY 10027\\
$^{6}$Department of Physics and Astronomy, University of North Carolina at Chapel Hill, Chapel Hill, NC 27599-3255, USA\\
$^{7}$European Southern Observatory, Karl-Schwarzschild-Str. 2, D-85748 Garching bei M\"{u}nchen, Germany\\
$^{8}$Department of Physics, University of Warwick, Gibbet Hill Road, Coventry, CV4 7AL, UK\\
$^{9}$Centre for Exoplanets and Habitability, University of Warwick, Gibbet Hill Road, Coventry, CV4 7AL, UK\\
$^{10}$Department of Astronomy, The University of Texas at Austin, Austin, TX 78712, USA \\
$^{11}$NASA Sagan Fellow \\
$^{12}$Leiden Observatory, Leiden University, Niels Bohrweg 2, NL-2333 RA Leiden, the Netherlands\\
$^{13}$Department of Earth and Planetary Sciences, Tokyo Institute of Technology, 2-12-1 Ookayama, Meguro-ku, Tokyo 152-8551, Japan \\
$^{14}$Institute for Astronomy, University of Hawai`i at M\={a}noa, Honolulu, HI 96822, USA \\
$^{15}$Amateur Astronomer, 7507 52nd Place NE, Marysville, WA, 98270 \\
$^{16}$NASA Ames Research Center, Moffett Blvd, Mountain View, CA 94035, USA \\ 
$^{17}$INAF - Osservatorio Astrofisico di Catania, via S. Sofia, 78, 95123 Catania, Italy \\ 
}

\pubyear{2018}

\begin{document}
\label{firstpage}
\pagerange{\pageref{firstpage}--\pageref{lastpage}}
\maketitle

% Abstract of the paper
\begin{abstract}
We describe \laurel{} and \hardey{}, two systems identified in the {\it K2} data whose light curves contain episodic drops in brightness with shapes and durations similar to those of the young ``dipper" stars, yet shallower by $\sim$1--2 orders of magnitude. These ``little dippers" have diverse profile shapes with durations of $\simeq$0.5--1.0 days and depths of $\simeq$0.1--1.0\% in flux; however, unlike most of the young dipper stars, these do not exhibit any detectable infrared excess indicative of protoplanetary disks, and our ground-based follow-up spectra lack any signatures of youth while indicating these objects as kinematically old. After ruling out instrumental and/or data processing artifacts as sources of the dimming events, we investigate possible astrophysical mechanisms based on the light curve and stellar properties. We argue that the little dippers are consistent with transits of star-grazing exocomets, and speculate that they are signposts of massive non-transiting exoplanets driving the close-approach orbits. 
\end{abstract}

% Select between one and six entries from the list of approved keywords.
% Don't make up new ones.

%%%%%%%%%%%%%%%%%%%%%%%%%%%%%%%%%%%%%%%%%%%%%%%%%%%
%%%%%%%%%%%%%%%%%%%% INTRODUCTION %%%%%%%%%%%%%%%%%
%%%%%%%%%%%%%%%%%%%%%%%%%%%%%%%%%%%%%%%%%%%%%%%%%%%

\section{Introduction}

The space-based {\it Kepler} mission \citep{Borucki2016} and its {\it K2} successor \citep{Howell2014} have provided ultra-precise time-series photometry for hundreds of thousands of nearby stars. Light curves from these missions have been used to identify thousands of close-in transiting exoplanets \citep[]{Batalha2013,Crossfield2016,Mann2017} and also study other types of circumstellar material around young stellar objects \cite[e.g.,][]{Ansdell2016a,Stauffer2017,Cody2018}, main-sequence stars \cite[e.g.,][]{Boyajian2016,Rappaport2018}, and even white dwarfs \cite[e.g.,][]{Vanderburg2015}. 

In particular, the so-called ``dipper'' stars are young ($\lesssim10$~Myr), K/M-type pre-main sequence stars that exhibit deep ($\gtrsim$10\%) and moderate-duration ($\sim$0.5--2.0~day) drops in brightness with diverse time profiles \citep[e.g., see Figure 4 in][]{Ansdell2016a} that can appear quasi-periodically or aperiodically \citep[e.g., see Figure 3 in][]{Ansdell2016a} as well as episodically \cite[e.g.,][]{Scaringi2016}. Although the first known dippers \citep[e.g., AA Tau;][]{Bouvier1999} were discovered from the ground, and later with the CoRoT and {\it Spitzer} space missions \citep{Alencar2010, MC2011, Cody2014}, {\it K2}'s survey of nearby star-forming regions has greatly expanded studies of these objects \citep{Ansdell2016a, Ansdell2016b, Scaringi2016, Bodman2017, Hedges2018, Cody2018}. The dipper phenomenon is thought to be due to transits of circumstellar dust, likely related to primordial circumstellar disks, as these objects nearly all have clear infrared excesses and often exhibit line emission related to accretion \citep{Ansdell2016a}. Moreover, simultaneous optical and near-infrared time-series photometry has shown that the dips can be shallower at longer wavelengths, consistent with extinction by optically thin dust \citep{MC2011,Cody2014,Schneider2018}. 

The unprecedented precision of {\it Kepler} has also enabled detection of very shallow ($\lesssim1$\%) flux dips in the light curves of two F2V main-sequence stars, KIC~3542116 and KIC~11084727 \citep{Rappaport2018}. These dimming events occur aperiodically and have shapes characteristic of trailing dust tails \citep[i.e, asymmetric shapes with steep ingresses and slower egresses;][]{Lecavelier1999}, thus have been explained in terms of  transits of remnant circumstellar planetesimals, namely ``exocomets" \citep{Rappaport2018}. KIC~3542116 exhibited six $\simeq$0.05--0.1\% dips in its {\it Kepler} light curve, each lasting $\simeq$0.5--1.0~days, while KIC~1108472 had a single similarly shaped transit. \cite{Rappaport2018} did not report any detectable infrared excess or other signatures of stellar youth for KIC~3542116 and KIC 11084727.  Similarly, the F3 V/IV star KIC~8462852 \cite[Boyajian's star;][]{Boyajian2016} lacks detectable infrared excess or signatures of youth, yet exhibits both shallow ($\sim$1\%) and deep ($\sim$20\%) flux drops with irregular shapes and typical durations of a few days. A variety of mechanisms have been invoked to explain KIC~8462852's dimming events, such as collisions of large bodies, a family of exocomet fragments, and a dusty debris ring \citep{Boyajian2016,Katz2017,Wyatt2018}. Regardless of the mechanism, the dips are likely caused by dusty material, as multi-band photometric monitoring from the ground has shown that the dips have a wavelength dependence consistent with extinction by optically thin sub-$\mu$m dust \citep{2018Bodman}.

Here we present two systems, \laurel{} and \hardey{}, which also do not exhibit any detectable infrared excesses or other signatures of youth, yet show very shallow ($\simeq$0.1--1\%) episodic dips in their {\it K2} light curves that last $\simeq$0.5--1.0~days. What sets these apart from the aforementioned exocomet systems is that most of the observed dips do {\em not} have the typical profiles of trailing dust tails; rather, the dip profiles have a variety of shapes (symmetric, leading-tail, complex) remarkably similar to those seen in the young dipper systems, but an order of magnitude shallower. In this work, we analyze the {\it K2} light curves and follow-up observations of these two ``little dippers." In Section~\ref{sec-data}, we present the available data for \laurel{} and \hardey{}, including their {\it K2} light curves and all-sky survey photometry, as well as our follow-up spectra and adaptive optics imaging. The stellar and dip properties of the little dippers are derived from these data in Section~\ref{sec-analysis}. We discuss possible mechanisms for the dimming events in Section~\ref{sec-discussion} and summarize our work in Section~\ref{sec-conclusion}.

%%%%%%%%%%%%%%%%%%%%%%%%%%%%%%%%%%%%%%%%%%%%%%%%%%%
%%%%%%%%%%%%%%%%%%%%%%% DATA %%%%%%%%%%%%%%%%%%%%%%
%%%%%%%%%%%%%%%%%%%%%%%%%%%%%%%%%%%%%%%%%%%%%%%%%%%

\section{Data}
\label{sec-data}

\subsection{K2 Light Curves}
\label{sec-k2}

\laurel{} and \hardey{} were discovered during a visual re-survey of {\it K2} lightcurves, focused on finding dipper stars that were outside of the constraints used in the original survey \cite[e.g., requiring at least five $>$10\% flux dips in the 80-day {\it K2} campaign;][]{Ansdell2016a} while also including data from newly released {\it K2} campaigns (up until Campaign 17). The search was conducted using {\tt LcTools},\footnote{https://sites.google.com/a/lctools.net/lctools/} a free and publicly available software program that provides a set of applications for efficiently building and visually inspecting large numbers of light curves \citep{Kipping2015}. For more details on the {\tt LcTools} package and the visual survey methodology, see \cite{Rappaport2018}. \laurel{} and \hardey{} were identified as distinct objects in the {\it K2} dataset based on their dipper-like profiles but very shallow transit depths. 

\subsubsection{EPIC~205718330}
\label{sec-k2-330}

\laurel{} was observed during {\it K2}~Campaign~2 (K2/C2). The 77.5-day K2/C2 light curve, shown in Figure~\ref{fig-lc1}, was extracted using the {\it K2} Self Field Flattening (K2SFF) technique described in \cite{Vanderburg2014} and \cite{Vanderburg2016}. K2SFF extracts light curves from {\it Kepler} Target Pixel Files (TPFs) using fixed photometric apertures, correcting for spacecraft motion by correlating observed flux variability with spacecraft pointing. This correction is needed because quasi-periodic thruster firings that account for spacecraft pointing drift can introduce artificial systematics into the {\it K2} light curves as differing amounts of target flux are lost or contaminated within the fixed photometric aperture.

\begin{figure*}
\begin{centering}
\includegraphics[width=17cm]{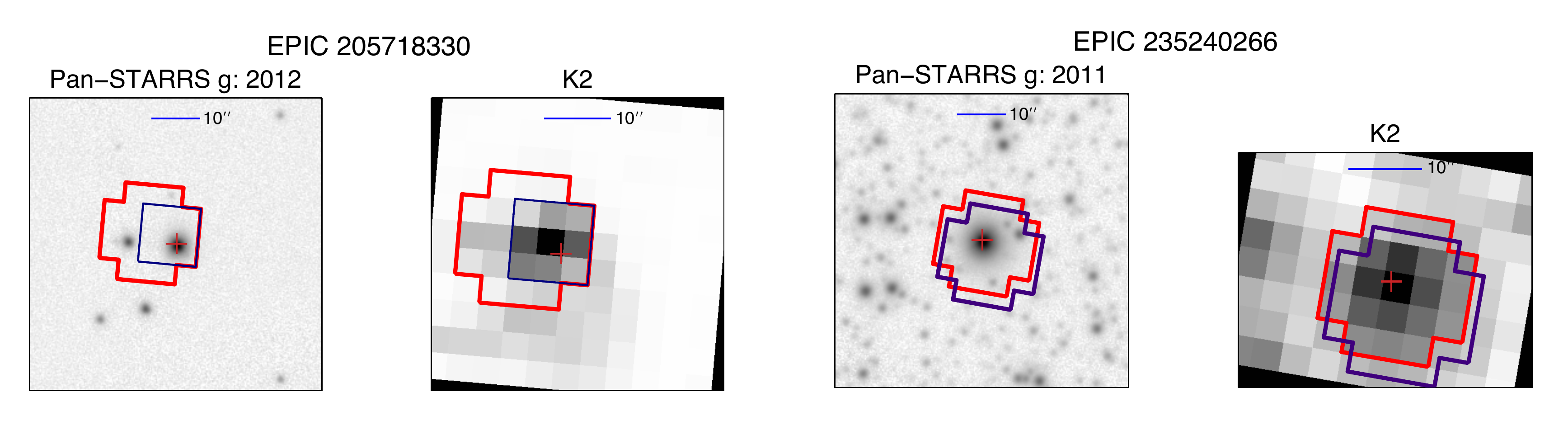}
\caption{\small Photometric apertures used to extract the light curves shown in Figures~\ref{fig-lc1} \& \ref{fig-lc2}, over-plotted on Pan-STARRS and {\it K2} images of \laurel{} (left) and \hardey{} (right). For \laurel{}, the red aperture was used to produce the light curve shown in Figure~\ref{fig-lc1}, while the blue aperture excludes two nearby stars but still produces a light curve that exhibits the dipping events (see Section~\ref{sec-k2-330}). For \hardey{}, the blue aperture was used to extract the light curve during the first half of K2/C11, while the red aperture was used to extract the light curve after \textit{Kepler}'s pointing was adjusted (see Section~\ref{sec-k2-266}).}
\label{fig-apertures}
\end{centering}
\end{figure*}

\begin{figure*}
\begin{centering}
\includegraphics[width=17cm]{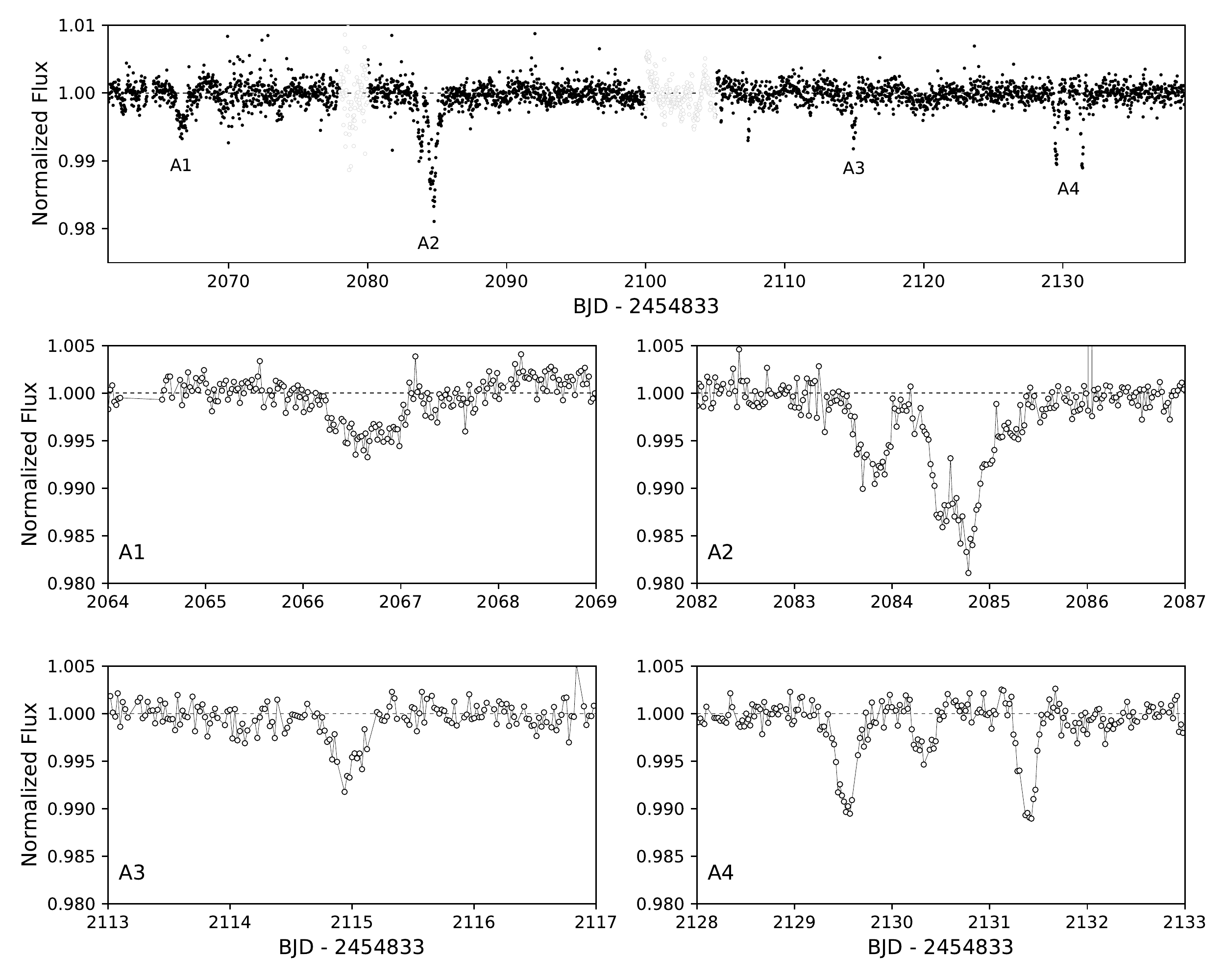}
\caption{\small {\it Top}: normalized K2/C2 light curve of \laurel{} (Section~\ref{sec-k2-330}). {\it Bottom}: closer looks at the four main dipping events.}
\label{fig-lc1}
\end{centering}
\end{figure*}

\begin{figure*}
\begin{centering}
\includegraphics[width=17cm]{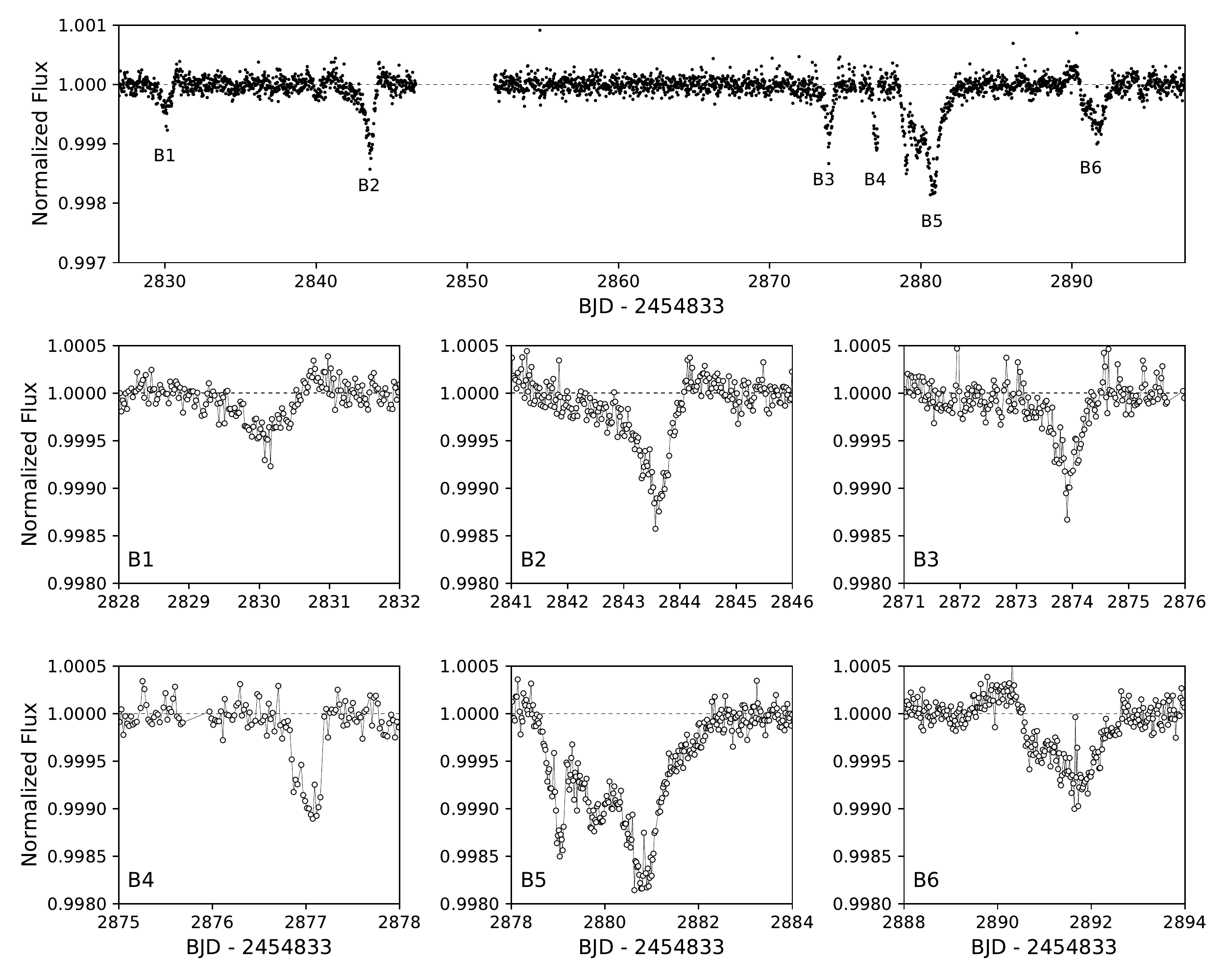}
\caption{\small {\it Top}: normalized K2/C11 light curve of EPIC~235240266 (Section~\ref{sec-k2-266}). {\it Bottom}: closer looks at the six main dipping events.}
\label{fig-lc2}
\end{centering}
\end{figure*}

Although the K2SFF light curves are made publicly available via the Mikulski Archive for Space Telescope (MAST),\footnote{https://archive.stsci.edu/prepds/k2sff/} we improved upon the default K2SFF output by performing the systematics correction on the full {\it K2} light curve, rather than splitting the time series into two segments; this eliminated a jump midway through the default K2SFF light curve. As shown in Figure \ref{fig-apertures}, we extracted the light curve using a large aperture to improve photometric precision, however one consequence was the inclusion of two contaminating stars; we confirmed that the dipping events were associated with \laurel{} by re-extracting the light curve using a smaller aperture that excluded these other stars, but also some flux from \laurel{}. The normalized light curve shown in Figure~\ref{fig-lc1} has been corrected for dilution due to the two contaminant stars by subtracting their fractional flux; for this, we used the {\it Gaia} Data Release 2 \cite[DR2;][]{Gaia2018} $G$-band magnitudes, which are similar to {\it Kepler} magnitudes (the dilution correction was ultimately negligible). To normalize the light curve, we divide by a fitted cubic spline with uniform knots every three days constructed using the {\tt LSQUnivariateSpline} interpolation in {\tt SciPy}. When fitting the spline, we excluded sections of the light curve containing the dips as well as two regions of increased scatter that are likely {\em not} due to astrophysical phenomenon associated with the target star (these are grayed out in Figure~\ref{fig-lc1}). We then interpolated the fitted spline over the excluded regions and applied the correction to the entire light curve.

\subsubsection{EPIC~235240266}
\label{sec-k2-266}

\hardey{} was targeted during {\it K2} Campaign~11 (K2/C11). K2/C11 was separated into two operational segments due to an error in the initial roll-angle used to minimize solar torque on the spacecraft. An excess roll motion identified at the beginning of K2/C11 indicated that targets would eventually move out of their set apertures. Therefore a $-0.32^\circ$ roll offset was applied 23 days into K2/C11, requiring new target aperture definitions, shown in Figure~\ref{fig-apertures}. The two light curve segments are identified separately in the MAST archive as C111 and C112. We downloaded the default K2SFF light curves, however due to the break in operations during K2/C11, there remained a jump in the data between the two segments as well as upward ``hooks'' at the beginning of each segment due to thermal settling of the spacecraft. Thus we removed the first two days of each segment before normalizing with fitted cubic splines with uniform knots every 1.5 and 1.0 days for the first and second segments, respectively. We note that the dilution corrections for the much fainter contaminant stars in the photometric aperture were negligible. Figure~\ref{fig-lc2} shows the full 70.5-day K2/C11 light curve.

\subsubsection{Validation Checks of K2 Light Curves}

We conducted validation checks to rule out common false-positive dip signals seen in {\it K2} data. First, we re-ran the K2SFF de-trending using a range of parameters to assess the robustness of the dips. Although the dips can be made slightly shallower or deeper depending on the chosen de-trending parameters, the changes were always less than a factor of two for reasonable values, and no combination of de-trending parameters could completely erase any of the dips. Thus we are confident that the dips are not a bi-product of the K2SFF de-trending process. Second, we checked for time-variable background noise (``rolling bands") and contamination from nearby ($\lesssim$5 arcmin) bright stars that could explain the dips. No time-variable background signals were seen around the targets and no similar dimming events were found in nearby stars on the same CCD module during the dips. Third, we tested for ``CCD crosstalk" that occurs when bright stars cause signals at the same pixel coordinates on other channels within the same module due to the coupled CCD readout. For both targets, we inspected all downloaded stars within a given radius on each channel in the module, but found no correlations between the dips and variability in nearby stars. To check against sources not downloaded individually, we inspected the full-frame images (FFIs) for nearby bright stars: only \hardey{} had a nearby saturated star on a different channel, however our target is outside of the halo and away from the bleed column, thus unlikely to be affected by crosstalk. Although \hardey{} could still be on a diffraction spike, there is no evidence of crosstalk on neighboring individually downloaded stars.

\subsection{Literature Data}
\label{sec-lit}

Literature data come from all-sky photometric surveys and {\it Gaia} DR2. Table~\ref{tab-prop} gives the precise coordinates ($\alpha_{\rm J2000}$, $\delta_{\rm J2000}$), proper motions ($\mu_{\alpha}$, $\mu_{\delta}$), and distances ($d$) from {\it Gaia} DR2. Table~\ref{tab-phot} gives the available photometry: optical photometry is from the AAVSO Photometric All Sky Survey \cite[APASS;][]{Henden2016} and {\it Gaia} DR2; near-infrared photometry is from the Deep Near Infrared Survey of the Southern Sky \citep[DENIS;][]{DENIS2005} and Two Micron All-Sky Survey \cite[2MASS;][]{Skrutskie2006}; mid-infrared photometry is from the Wide-field Infrared Survey Explorer \cite[WISE;][]{Wright2010}. The 2MASS designations for \laurel{} and \hardey{} are 2MASS~16333538-1530414 and 2MASS~17241057-2332318, respectively.

\bgroup
\def\arraystretch{1.5}
\setlength{\tabcolsep}{.2em}
\begin{table}
\caption{Stellar Properties}
\begin{tabular}{lrrl}
\hline \hline
Parameter & 205718330 & 235240266 & Units  \\
\hline \hline
{\it Gaia} \\
\hline
$\alpha_{\rm J2000}$    & 16:33:35.3702          & 17:24:10.5454      & \\
$\delta_{\rm J2000}$     & $-15$:30:42.420       & $-23$:32:32.636    & \\
$\mu_{\alpha}$              & $-14.867\pm0.070$  & $-21.722\pm0.102$  & mas/yr \\
$\mu_{\delta}$              & $-55.413\pm0.047$   & $-46.900\pm0.081$  & mas/yr \\
$d$                               & $240.7\pm2.6$          & $334.6\pm7.5$      & pc \\
\hline 
Photometry + Parallax \\
\hline
$T_{\rm eff}$        &  4810    & 6120     & K \\
${A_V}$                &  1.84    & 1.60      & mag \\
log~$g$                &  4.57    & 4.10      &   \\
${\rm [Fe/H]}$       &  0.13   & 0.04      &  \\
${\rm R_{\star}}$  &  0.76    & 1.64      & ${\rm R_{\odot}}$ \\
${\rm M_{\star}}$  &  0.78   & 1.20      & ${\rm M_{\odot}}$ \\
\hline
X-Shooter Spectra \\
\hline
$T_{\rm eff}$      &  $4850\pm200$           &    $5850\pm120$             & K \\
$\log g$              &  $3.7\pm0.5$               &    $3.2\pm0.4$                 &   \\
$v \sin i$             & $<8$                           &     $11 \pm2$              & km~s$^{-1}$ \\
RV                      & $-12.8\pm2.0$            &     $-28.5\pm1.0$             & km~s$^{-1}$ \\
\hline
SNIFS + SpeX Spectra \\
\hline
$T_{\rm eff}$      &  4900           & 5700             & K \\
$A_V$                &  2.1              & 1.4                & mag \\
log~$g$              &  4.0              & 3.5                & \\
${\rm [Fe/H]}$     & $-0.5$         & $-0.5$           &  \\
\hline
\hline
\end{tabular}
\label{tab-prop}
\end{table}

%Star                       Teff    err      logg   err      vsini   err    RV  err
%EPIC205718330       4650  250     4.0    0.5      <8             -12.8  2.0
%EPIC235240266       5850  150     3.5    0.5      11     2      -28.5  1.0

\bgroup
\def\arraystretch{1.5}
\setlength{\tabcolsep}{1.3em}
\begin{table}
\caption{Photometry}
\begin{tabular}{lll}
\hline \hline
Band & EPIC~205718330 & EPIC~235240266  \\
\hline \hline
Johnson $B$        &  $16.767\pm0.080$   & $13.670\pm0.040$ \\
Johnson $V$        &  $15.209\pm0.048$   & $12.648\pm0.031$ \\
Sloan $g'$         &  $16.056\pm0.040$   & $13.134\pm0.031$ \\
Sloan $r'$         &  $14.553\pm0.030$   & $12.303\pm0.040$ \\
Sloan $i'$         &  $14.144\pm0.360$   & $11.908\pm0.060$ \\
$G_{\rm BP}$       &  $15.501\pm0.001$   & $12.919\pm0.002$ \\ 
$G_{\rm G}$        &  $14.523\pm0.002$   & $12.298\pm0.001$ \\
$G_{\rm RP}$       &  $13.549\pm0.001$   & $11.554\pm0.002$ \\ 
2MASS $J$          &  $12.095\pm0.027$   & $10.494\pm0.022$ \\
2MASS $H$          &  $11.389\pm0.024$   & $10.039\pm0.022$ \\
2MASS $K_{\rm S}$  &  $11.177\pm0.019$   & $9.911\pm0.023$ \\
WISE~1             &  $11.054\pm0.024$   & $9.831\pm0.023$ \\
WISE~2             &  $11.099\pm0.028$   & $9.898\pm0.021$ \\
WISE~3             &  $10.984\pm0.045$   & $9.935\pm0.099$ \\
WISE~4             &  $> 8.331$          & $>8.458$       \\
\hline \hline
\end{tabular}
\label{tab-phot}
\end{table}

\subsection{Follow-up Observations}

\subsubsection{Spectroscopy}
\label{sec-spec}

We acquired spectra with the wide-band, intermediate-resolution X-Shooter spectrograph \citep{Vernet2011} mounted on the 8.2~m European Southern Observatory (ESO) Very Large Telescope (VLT) at Cerro Paranal in Chile during UT 2018 May 20--21. VLT/X-Shooter simultaneously covers wavelengths from about 300 to 2500~nm divided into UVB (300--550~nm), VIS (500--1050~nm), and NIR (1000--2500~nm) arms. The slit widths differ for each arm: we used narrow slits (0.1\as{}, 0.4\as{}, 0.4\as{}) to obtain finer spectral resolution ($R\simeq5400$, 18400, 11600) in addition to wide slits (5.0\as{}, 5.0\as{}, 5.0\as{}) that do not suffer from flux losses for absolute flux calibration. Data reduction was performed with the ESO X-Shooter pipeline \citep{Modigliani2010} version 2.9.3, which includes flat fielding, bias subtraction, order extraction and combination, rectification, wavelength calibration, flux calibration using standard stars observed in the same night, and final extraction of the spectrum. 

We also obtained optical spectra using the moderate-resolution Super-Nova Integral Field Spectrograph \citep[SNIFS;][]{Aldering2002,Lantz2004} at the University of Hawaii 2.2 m telescope atop Maunakea during UT 2018 May 18--19. SNIFS covers wavelengths from 3200 to 9700~\AA{} and does not suffer from wavelength-dependent slit losses that can be difficult to accurately correct. Our SNIFS spectra have resolutions of $R\simeq900$ and signal-to-noise ratios of ${\rm SNR}\simeq100$ per resolution element at 6500 \AA{}. Details of our SNIFS observations, data reduction, and extraction can be found in \cite{Mann2012} and \cite{Lepine2013}. 

We acquired moderate-resolution near-infrared spectra using the upgraded SpeX spectrograph \citep[Spex;][]{Rayner2003} on the 3.2~m NASA Infrared Telescope Facility (IRTF) atop Maunakea on UT 2018 April 24. Our SpeX spectra were taken in the short cross-dispersed (SXD) mode using the 0.3\as{}$\times$15\as{} slit, covering 0.7 to 2.5 $\mu$m at $R\approx2000$ with ${\rm SNR}\gtrsim80$ in $K$ band. Basic reduction (bias subtraction, flat fielding, extraction, etc.) was carried out with SpeXTool \citep{Cushing2004}. Flux calibration and telluric line removal were then performed using A0V standards with Xtellcor \citep{Vacca2003}. See \cite{Mann2013} for details on the observations, data reduction, and spectrum extraction.

\subsubsection{AO Imaging}
\label{sec-ao}

Adaptive optics (AO) imaging for \laurel{} was acquired using the Infrared Camera and Spectrograph \citep[IRCS;][]{Kobayashi2000} with AO88 \citep{Hayano2010} on the Subaru Telescope atop Maunakea on UT 2018 June 14. We performed $K^\prime$ imaging in fine-sampling mode using a five-point dither pattern with a total integration of 37.5~s. Weather conditions were good and natural seeing was $0\farcs4-0\farcs6$ in near-infrared bands. We reduced the raw IRCS data using the procedures of \cite{Hirano2016}, which include dark subtraction, flat fielding, and distortion correction of the individual frames, which were then aligned and median combined to form the final image. For \hardey{}, we used the Near InfraRed imaging Camera (NIRC2) on the Keck II 10 m telescope atop Maunakea with natural guide star AO \cite[NGS-AO;][]{Wizinowich2000,vanDam2004} on UT 2018 April 29. Seeing conditions were poor and only partial corrections were obtained in most images: the best resolution was $0\farcs6$. Eight sets of 10$\times$0.5 integrations were obtained through the $K^{\prime}$ filter with the narrow camera.  

\subsubsection{Long-baseline Light Curves}
\label{sec-asassn}

To assess long-term ($\sim$yrs) variability, we obtained photometry from the All-Sky Automated Survey for SuperNovae \cite[ASAS-SN;][]{Shappee2014,Kochanek2017}. ASAS-SN images the sky every two days down to $V\sim17$ from CTIO in Chile and Haleakala in Hawaii, both hosted by the Las Cumbres Observatory Global Telescope Network \citep[LCOGT;][]{Brown2013}. Each site has four 14~cm lenses, each with a 2k$\times$2k CCD camera. The field of view is 4.5$\times$4.5~deg, the pixel scale is 8\as{}, and the FWHM of the PSF is 15\as{}. The $V$-band magnitudes for each source were extracted from the images using aperture photometry with zero-points calibrated using the APASS catalog \citep{Levine2017}. \laurel{} was observed $\simeq$1000 times from UT 2013 Feb 14 to 2018 June 3 with a median per-point error of 0.06~mags, while \hardey{} was observed $\simeq$600 times from UT 2015 February 16 to 2018 May 22 with a median per-point error of 0.02~mags. The ASAS-SN light curves for both stars are shown in Figure~\ref{fig-asassn}.

\begin{figure}
\includegraphics[width=8.5cm]{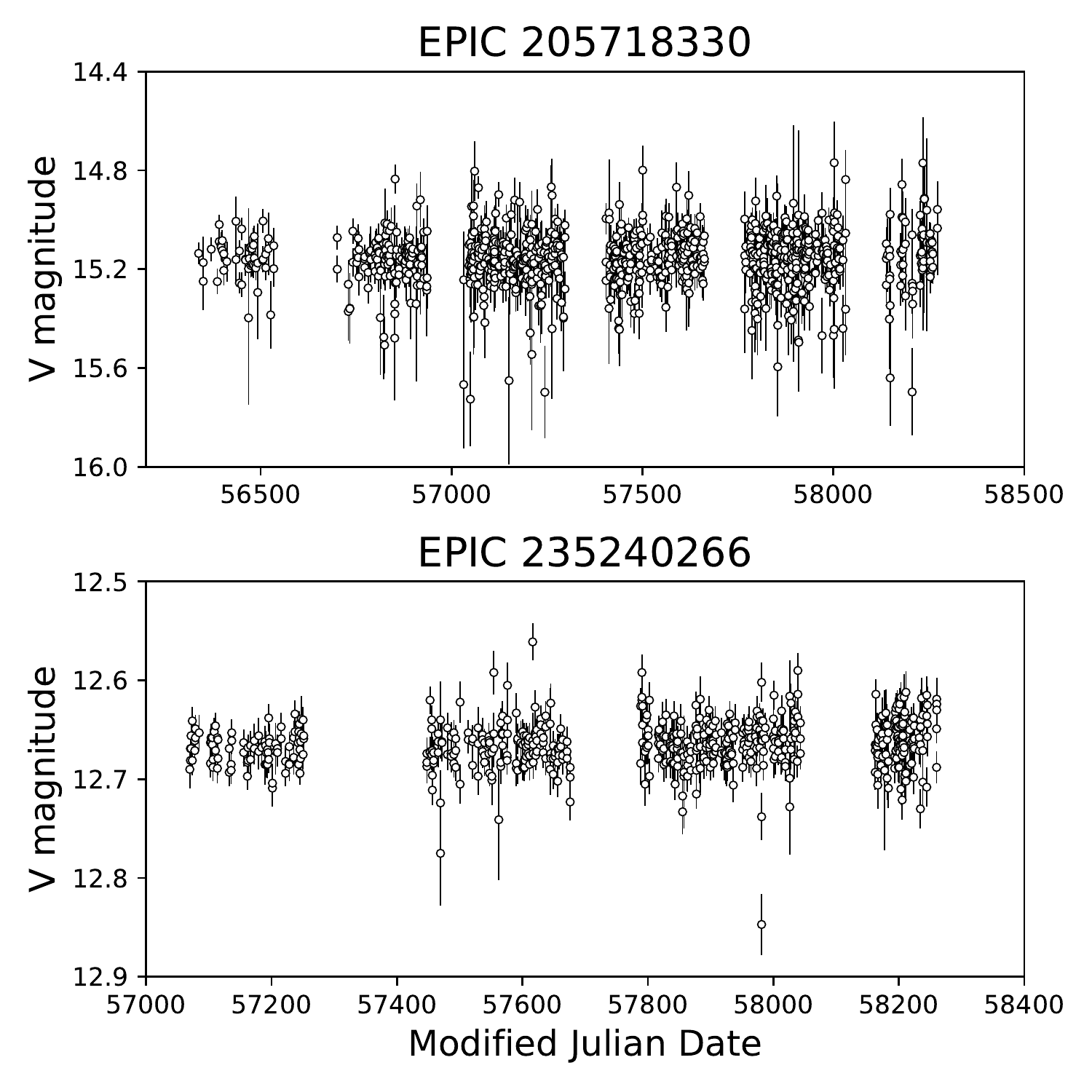}
\caption{\small Long-baseline light curves from ASAS-SN (Section~\ref{sec-asassn}), spanning 5.3 years for \laurel{} and 3.3 years for \hardey{}. No long-term trends are evident, and the single-point drops in flux typically have uncertainties $\gtrsim2\times$ the median error thus are likely unreliable measurements.}
\label{fig-asassn}
\end{figure}

%%%%%%%%%%%%%%%%%%%%%%%%%%%%%%%%%%%%%%%%%%%%%%%%%%%
%%%%%%%%%%%%%%%%%%%%% ANALYSIS %%%%%%%%%%%%%%%%%%%%
%%%%%%%%%%%%%%%%%%%%%%%%%%%%%%%%%%%%%%%%%%%%%%%%%%%

\section{Analysis}
\label{sec-analysis}

\subsection{Stellar Properties}
\label{sec-stellar}

\subsubsection{Photometrically derived properties}
\label{sec:starphot}

We fit the {\it Gaia} parallax in Table~\ref{tab-prop} and the observed photometry in Table~\ref{tab-phot} (except for the {\it Gaia} magnitudes) to the MESA Isochrones \& Stellar Tracks \cite[MIST;][]{Dotter2016} model grid using the {\tt isochrones} Python package \citep{Morton2015}. {\tt isochrones} is a tool for inferring model-based physical properties given photometric or spectroscopic observations of a star. The package performs 3D linear interpolations in mass--metallicity--age parameter space across a given stellar model grid and then uses nested sampling to capture the potentially multi-modal posterior distributions of the stellar physical parameters (e.g., in the case of evolved stars along the subgiant branch). For more details on the fitting method and applications to other systems, see \citet{Montet2015} and \citet{Morton2016}. The best-fit results are shown in Table~\ref{tab-prop}; typical errors were a few percent, although these are likely underestimated. Nevertheless, the results indicate that \laurel{} and \hardey{} have stellar effective temperatures ($T_{\rm eff}$) of $\approx$5000~K and $\approx$6000~K, corresponding to main sequence mid-K and late-F dwarf stars, respectively. They also both have metallicities ([Fe/H]) consistent with solar and moderate extinctions ($A_V$), as expected for their locations. We note that \hardey{} has a surface gravity of $\log g \approx 4.0$, indicating it may be ascending the sub-giant branch, which is also suggested by its $K_{\rm S}$ magnitude and {\it Gaia} distance.

\subsubsection{Spectroscopically Derived Properties}
\label{sec:starspec}

We used our intermediate-resolution X-Shooter spectra (Section~\ref{sec-spec}) to derive the stellar properties of \laurel{} and \hardey{} given in Table~\ref{tab-prop}. For this, we utilized {\tt ROTFIT} \citep{Frasca2017}, which fits BT-SETTL synthetic stellar photosphere templates \citep{Allard2011} to several segments of continuum-normalized X-Shooter spectra in order to derive atmospheric parameters ($T_{\rm eff}$ and $\log g$) as well as projected rotational velocity ($v \sin i$) by $\chi^2$ minimization. We assumed solar metallicity and zero veiling (note that because {\tt ROTFIT} does not fit the continuum shape, assumptions of extinction do not affect the results). For \laurel{} the best-fit model had $T_{\rm eff} = 4850\pm200$~K and $\log g = 3.7\pm0.5$, while for \hardey{} the best-fit model had $T_{\rm eff} = 5850\pm120$~K and $\log g = 3.2\pm0.6$---all roughly consistent with the values derived from photometry and {\it Gaia} parallax (Section~\ref{sec:starphot}). We also find low $v \sin i$ values of $< 8$~km~s$^{-1}$ for \laurel{} and $11 \pm2$~km~s$^{-1}$ for \hardey{} (as shown in \citealt{Frasca2017}, the resolution and sampling of X-Shooter spectra do not allow constraints on $v \sin i$ lower than 6 or 8 km~s$^{-1}$, depending on the slit width). {\tt ROTFIT} also measures the heliocentric radial velocity (RV) by means of a Gaussian fit to the cross-correlation function between the target and synthetic spectrum, where the heliocentric correction is calculated by the X-Shooter pipeline; we found RV$=-12.8\pm2.0$~km~s$^{-1}$ for \laurel{} and RV$=-28.5\pm1.0$~km~s$^{-1}$ for \hardey{}. 

\begin{figure}
\includegraphics[width=7.2cm]{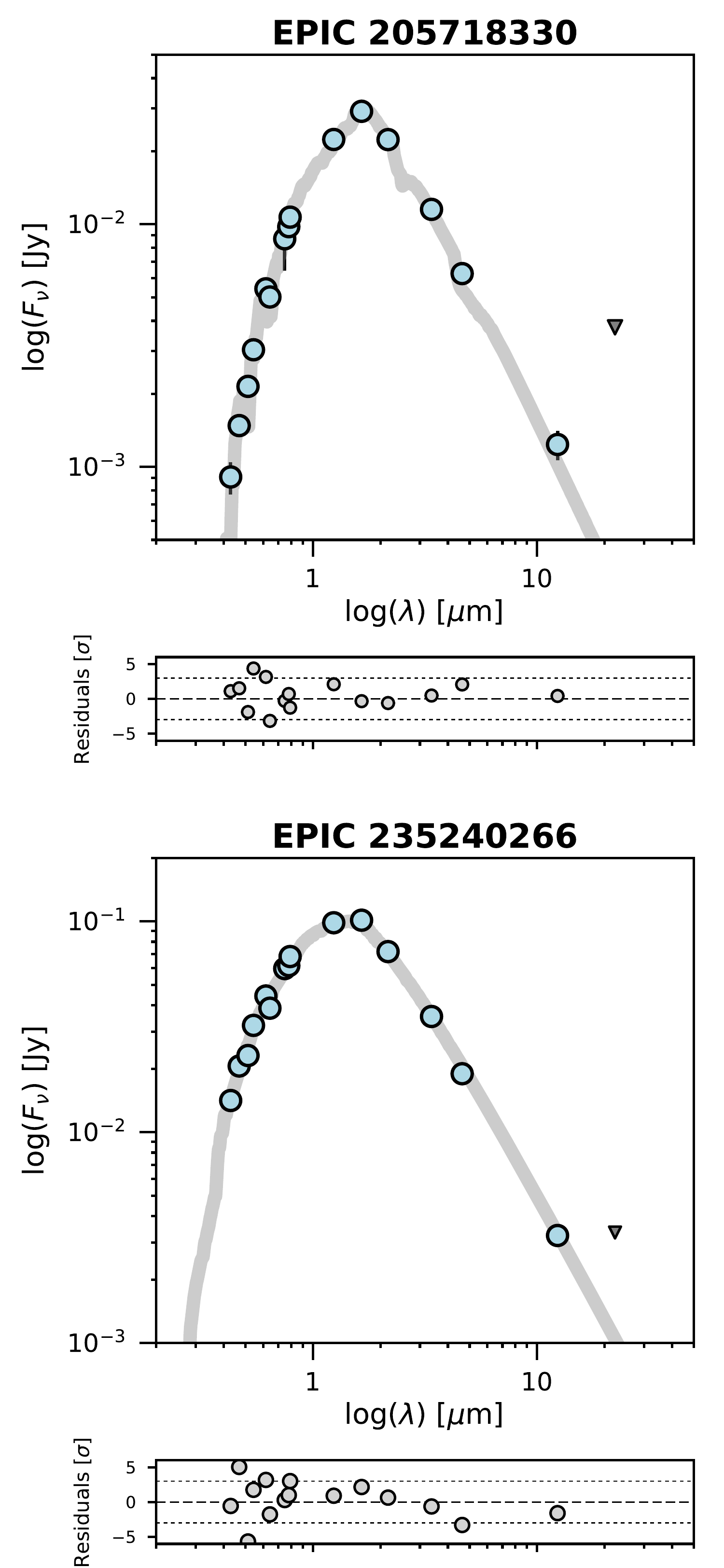}
\caption{\small SEDs of  \laurel{} (top) and  \hardey{} (bottom) using photometry from Table~\ref{tab-phot}. Blue circles are detections, gray triangles are upper limits, and error bars are typically smaller than the symbols. Thick gray lines are best-fit PHOENIX stellar photosphere models (Section~\ref{sec-excess}). Sub-panels show residuals of the observed and synthetic photometry in units of measurement uncertainty ($\sigma$); dotted horizontal lines denote $\pm3\sigma$ limits. The SEDs illustrate the lack of significant infrared excess above the stellar photosphere for both targets.}
\label{fig-sed}
\end{figure}

Additionally, we used our moderate-resolution visible SNIFS spectra and near-infrared SpeX spectra (Section~\ref{sec-spec}) to check these stellar properties while varying [Fe/H] and $A_V$. We prepared a single flux-calibrated spectrum for each target by splicing their SNIFS and SpeX spectra together, slightly distorting the spectra to achieve minimum $\chi^2$ agreement between synthetic magnitudes generated from the observed spectra using the profiles determined in \citet{Mann2015b} and the actual APASS and 2MASS photometry. We then compared these flux-calibrated spectra to a grid of model atmosphere spectra generated by the PHOENIX code \citep{Husser2013}, with steps of 100~K in $T_{\rm eff}$, 0.5 in $\log g$, and 0.5 in [Fe/H]. For each comparison model spectrum the best-fit $A_V$ was calculated, adopting the \citet{Cardelli1989} extinction model. These stellar properties are given in Table~\ref{tab-prop}; we do not give uncertainties due to strong degeneracies between $T_{\rm eff}$ and $A_V$. For \laurel{}, the best-fit model had $T_{\rm eff} \approx 4900$~K, $A_V \approx 2.07$, $\log g \approx 4.0$, and ${\rm [Fe/H]} \approx -0.5$---all roughly consistent with the values derived from the photometry and {\it Gaia} parallax (Section~\ref{sec:starphot}; unsurprising since the spectra are flux-calibrated with photometry) as well as our X-Shooter spectra (see above).  The expected total extinction along this line of sight is $A_V\approx2.2$ \citep{Schlegel1998, Schlafly2011}, which could be consistent with our best-fit value if most of this total extinction is due to the foreground Upper Scorpius cloud. For \hardey{}, the best fit model had $T_{\rm eff} \approx 5700$~K, $A_V \approx1.4$, $\log g \approx 3.5$, and ${\rm [Fe/H]} \approx -0.5$; this is again broadly consistent with the photometry and {\it Gaia} parallax as well as our X-Shooter spectra, and are allowed by the expected total extinction along this line of sight of $A_V\approx3.2$.

Stellar properties based on photometry and the overall spectral energy distribution (SED) shape are subject to systematic errors produced by interstellar reddening, if $A_V$ is not independently determined, and to a lesser extent by the covariance between $T_{\rm eff}$, $\log g$, and [Fe/H].  This is particularly problematic for \hardey{}, as its photometry and parallax is consistent either with an F-type dwarf that is more reddened, or a K-type subgiant that is less reddened. These degeneracies have less impact on analyses based on spectra that resolve individual lines.  In particular, the narrowness of strong lines in the X-shooter spectrum of \hardey{} suggest lower $\log g$, but final adjudication will probably require a higher-quality spectrum than what is currently available.

% Finally, we used our X-Shooter spectra to obtain constraints on the heliocentric radial velocities (RV) of these stars. We analyzed 10 spectral lines in the optical and near-infrared, fitting Gaussians to each line and calculating the offset in velocity space from the expected line center. This included a correction for heliocentric motion calculated by the X-Shooter pipeline. We then took the medians and standard deviations of the offsets as our final estimates and uncertainties, finding RV$=-12\pm2$~km~s$^{-1}$ for \laurel{} and RV$=-29\pm2$~km~s$^{-1}$ for \hardey{}. 

\subsection{Signatures of Youth}
\label{sec-youth}

\subsubsection{Infrared Excess}
\label{sec-excess}

We used the observed photometry in Table~\ref{tab-phot} to construct SEDs, which we compared to synthetic photometry derived from PHOENIX stellar atmosphere models to constrain the levels of any infrared excess above the stellar photosphere that could be indicative of dusty circumstellar disks. We found the best-fit stellar model by minimizing the differences between the observed and synthetic photometry while varying $T_{\rm eff}$, $A_{\rm V}$, $\log g$, and [M/H]. The SEDs and best-fit models are shown in Figure~\ref{fig-sed}, which illustrate a lack of detectable infrared excess for both targets. The corresponding limits on the fractional luminosity $f = L_{\rm disk}/L_\star$ for disks approximated as blackbodies with temperatures of 300~K are $f<2\times10^{-3}$ and $f<3\times10^{-4}$ for \laurel{} and \hardey{}, respectively (limits at all other temperatures are higher). These modest limits are insufficient to rule out possible debris disks, which commonly have $f<1\times10^{-4}$, especially at ages older than $\sim$100~Myr \citep{Wyatt2008}. However, we can use these limits on infrared excess to clearly rule out primordial circumstellar disks, which sets these systems apart from the young dipper stars, as illustrated in Figure~\ref{fig-colors}. Additionally, the long-baseline light curves from ASAS-SN (Figure~\ref{fig-asassn}; Section~\ref{sec-asassn}) do not exhibit any long-term dimming that could indicate the presence of circumstellar material in the outer disk \cite[e.g.,][]{Rodriguez2017}.

\subsubsection{Spectroscopic Accretion Signatures}
\label{sec-acc}

Young objects hosting circumstellar disks exhibit spectral accretion signatures that are produced by shocked gas free-falling onto the star along magnetic field lines as well as disk winds that can be emitted via a variety of mechanisms. We therefore searched our X-Shooter spectra for optical and near-infrared magnetospheric accretion signatures, namely H$\alpha$ (6563 \AA{}), Pa$\gamma$ (1.094~$\mu$m), Pa$\beta$ (1.280~$\mu$m), and Br$\gamma$ (2.166~$\mu$m) emission from H {\small \sc I}. These emission lines are routinely observed with X-Shooter in accreting young stars \cite[e.g.,][]{Alcala2014,Alcala2017}. We also searched for the He {\small \sc I} (1.083 $\mu$m) line, which is particularly sensitive to inner disk flows; the line can exhibit red-shifted absorption due to in-falling gas and/or blue-shifted absorption due to inner disk winds, both of which can be shifted on the order of hundreds of km~s$^{-1}$ \citep[e.g.,][]{Edwards2006}. 

As shown in Figure~\ref{fig-lines}, neither \laurel{} nor \hardey{} show red- or blue-shifted He {\small \sc I} absorption and neither source shows H$\alpha$, Pa$\gamma$, Pa$\beta$, or Br$\gamma$ emission. Rather, all lines are seen in absorption at the expected line position or simply not detected. This behavior is notably different than what is seen for the young dipper stars, which often show these spectroscopic signatures of youth \citep{Ansdell2016a}, albeit more weakly than strongly accreting classical T Tauri stars \cite[e.g.,][]{Edwards2006}.

\begin{figure}
\includegraphics[width=8.3cm]{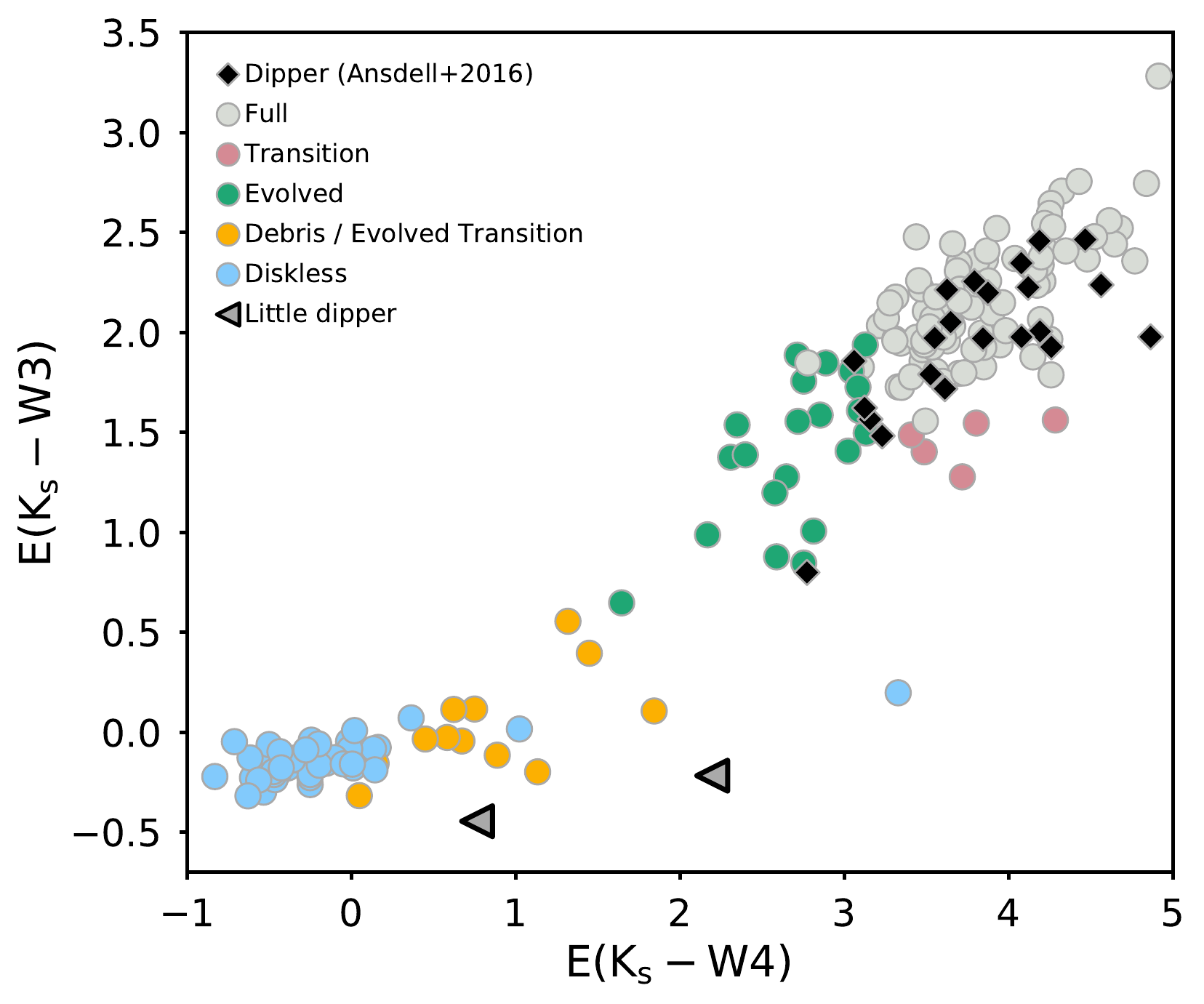}
\caption{\small Extinction-corrected WISE infrared color excesses used to classify disk types. Circles indicate late-type (K/M dwarf) Upper Sco members and colors specify their disk types; see \citet{LM2012} for details on the extinction corrections and disk classifications. The black diamonds are known dipper stars from \citet{Ansdell2016a} and the little dippers from this work are shown by the gray triangles, indicating upper limits on the WISE-4 excesses. The little dippers are clearly distinct from primordial disks and thus the dipper population, however we cannot rule out debris disks based on current limits (Section~\ref{sec-excess}).}
\label{fig-colors}
\end{figure}

\begin{figure*}
\begin{centering}
\includegraphics[width=17cm]{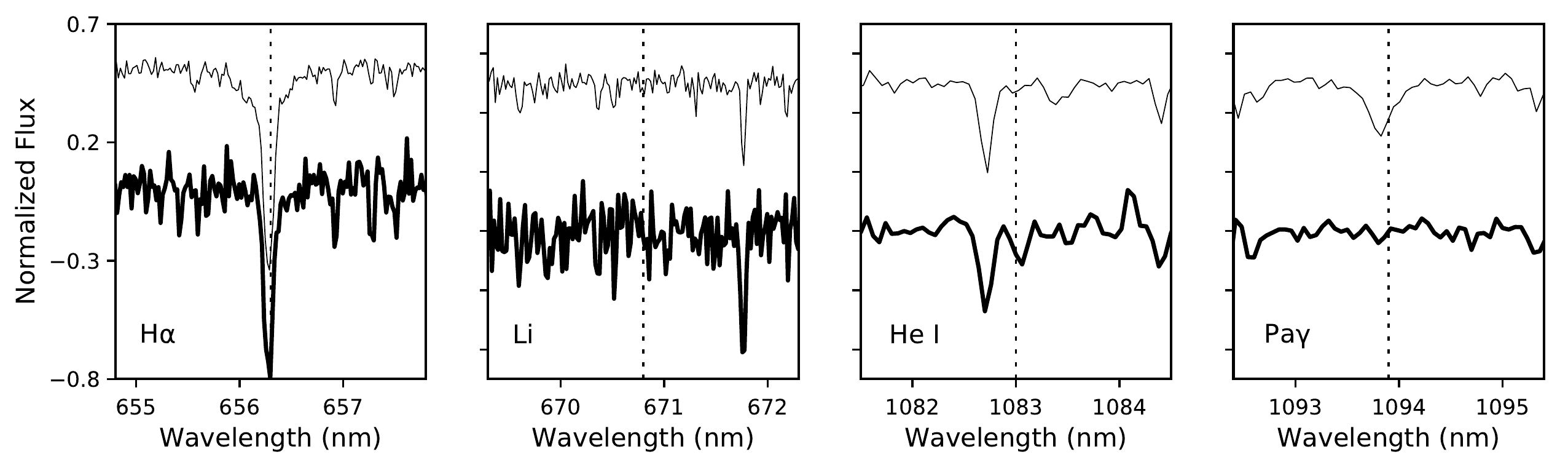}
\caption{\small Spectral lines used to assess stellar youth (Section~\ref{sec-acc}). Bold lines correspond to \laurel{}, while thin lines offset upward for clarity correspond to \hardey{}. Neither source shows detectable spectral signatures of stellar youth.}
\label{fig-lines}
\end{centering}
\end{figure*}

\subsubsection{Lithium Absorption}
\label{sec-li}

Another spectroscopic indicator of youth for low-mass stars is Li {\sc I} (6708.0 \AA{}) absorption. This is because low-mass stars have convective outer envelopes, which transport Li {\sc I} in their photospheres down into their hotter stellar cores where the element is destroyed. This process is rapid ($\lesssim50$~Myr) for mid-to-late M dwarfs with fully convective envelopes, but is slower for early-M and K dwarfs, as illustrated in Figure~\ref{fig-li}. 

X-Shooter is routinely used to detect Li {\sc I} absorption in young stars \cite[e.g.,][]{Manara2017b}. As shown in Figure~\ref{fig-lines}, our X-Shooter spectra show no signs of significant Li~{\sc I} absorption for either target: we find equivalent widths of EW$_{\rm Li} = -0.07\pm0.05$ for \laurel{} and EW$_{\rm Li} = 0.00\pm0.02$~\AA{} for \hardey{}. Uncertainties were determined using a Monte Carlo method: we used the standard deviation of the continuum regions around the expected line position to add Gaussian-distributed noise to the observed spectrum, then repeated the EW$_{\rm Li}$ measurement 100 times, taking the mean and standard deviation as our final EW$_{\rm Li}$ values and uncertainties, respectively. 

The lack of Li absorption allows us to place lower limits on stellar ages. Figure~\ref{fig-li} compares our 3$\sigma$ upper limits on EW$_{\rm Li}$ for both stars to the EW$_{\rm Li}$ values for members of young clusters of various ages as a function of $T_{\rm eff}$. The EW$_{\rm Li}$ values come from: \cite{Cummings2017} for the Hyades and Praesepe; \cite{Bouvier2018} for the  Pleiades; and \cite{daSilva2009} for all other regions. The rough ages given in Figure~\ref{fig-li} are taken from \cite{Gagne2018}. From Figure~\ref{fig-li}, we can constrain the age of \hardey{} to $\gtrsim800$~Myr. Due to the larger uncertainties for \laurel{}, we can only constrain its age to $\gtrsim150$~Myr. Note that these limits assume solar metallicity; for metal poor stars, the limits on age could be older.

\begin{figure}
\includegraphics[width=8.2cm]{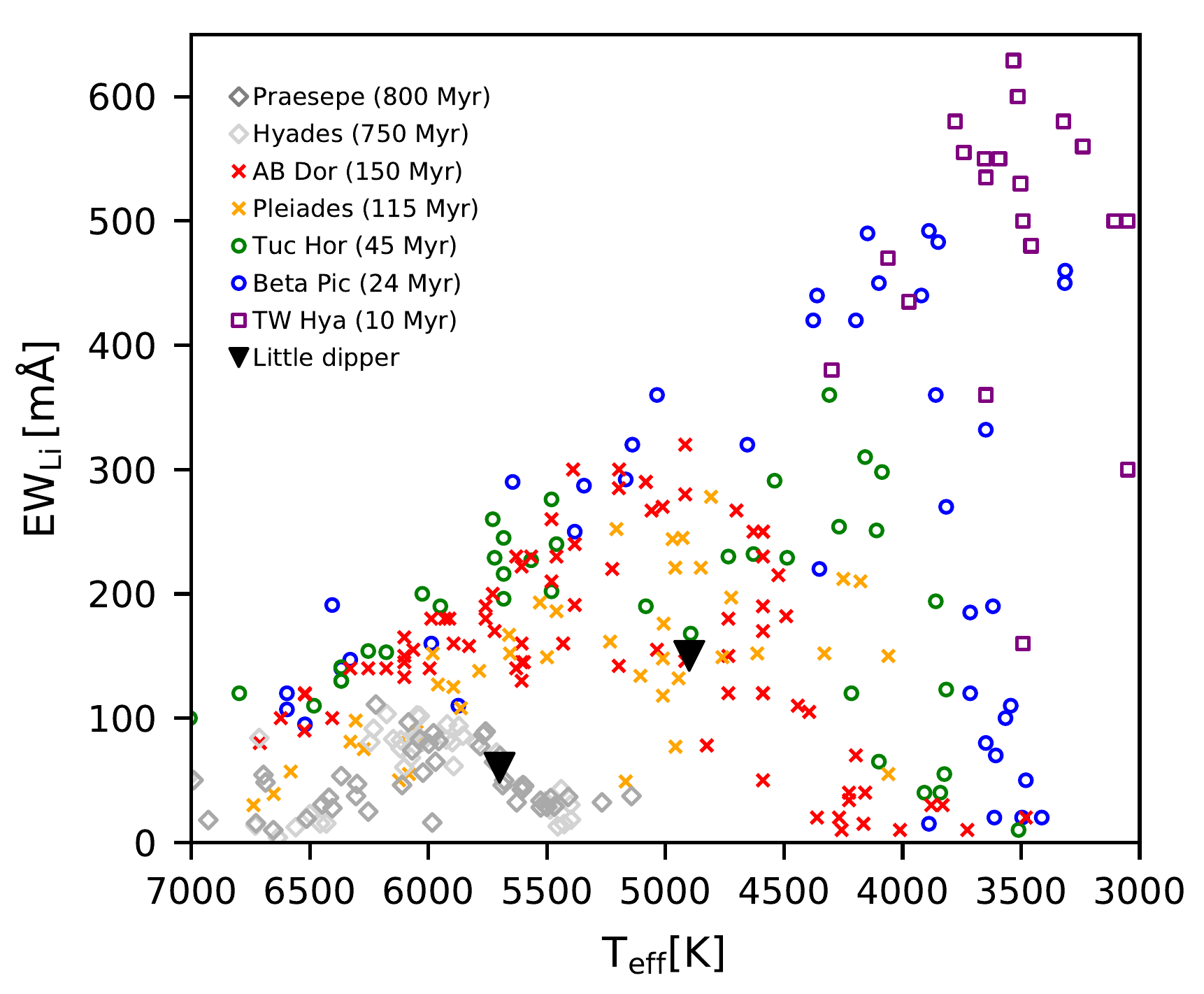}
\caption{\small Lithium equivalent widths (EW$_{\rm Li}$) as a function of stellar effective temperature ($T_{\rm eff}$) for young clusters of various ages, compared to our 3$\sigma$ EW$_{\rm Li}$ upper limits for the little dippers, illustrating that they are both likely $\gtrsim$150~Myr old.}
\label{fig-li}
\end{figure}

\subsubsection{Kinematics}
\label{sec-kinematics}

The strong constraints on parallax and proper motion from {\it Gaia} DR2 (Section~\ref{sec-lit}; Table~\ref{tab-prop}), combined with the RVs from our X-Shooter spectra (Section~\ref{sec:starspec}), can be used to determine whether \laurel{} or \hardey{} are associated with any young stellar populations that are still sufficiently bound to occupy distinct regions in both physical and kinematic space. In particular, young disk stars are known to cluster in a distinct ``box" of galactic space motion ($UVW$; calculated with respect to the Sun, where $U$ is positive toward the Galactic center) defined by $-20<U<50$, $-30<V<0$, and $-25<W<10$~km~s$^{-1}$ \citep{Leggett1992}.

% used this: http://www.astro.ucla.edu/~drodrigu/UVWCalc.html#data
% based on this: https://idlastro.gsfc.nasa.gov/ftp/pro/astro/gal_uvw.pro

For \laurel{}, we calculated Galactic space motions of $U=-0.36$, $V=-60.28$, and $W=-28.21$~km~s$^{-1}$. These values put it outside of the young disk parameter range due to the $V$ and $W$ space motion components. Indeed, although its proper motion of $\mu_{\alpha} = -15$~mas~yr$^{-1}$ and $\mu_{\delta}= -55$~mas~yr$^{-1}$ is consistent with the spectroscopically confirmed cool members of the $\sim$10~Myr old Upper Sco association \citep[e.g., see Figure~3 in][]{Bouy2009}, its distance of 240~pc puts it well beyond that of the region \citep[145~pc;][]{deZeeuw1999}. Moreover, its RV of $-12$~km~s$^{-1}$ is below the typical values for Upper Sco members, which have a median value and 1$\sigma$ velocity dispersion of $-6.31\pm4.61$~km~s$^{-1}$ for F2--K9 stars and $-6.28\pm3.04$~km~s$^{-1}$ for M0--M8 stars \citep{Dahm2012}. For \hardey{}, we calculated Galactic space motions of $U=-24.48$, $V=-81.90$, and $W=-15.97$~km~s$^{-1}$. This again puts it outside of the young disk parameter range due to its $U$ and $V$ space motion components.

\subsection{Nearby Sources or Companions}
\label{sec-nearby}

Our AO imaging of \laurel{} (Section~\ref{sec-ao}) did not show any nearby bright companions within 5\as. Our detection thresholds have contrast ratios of $\Delta K^\prime\approx7$~mags at separations of $\rho\approx1$\as{} or $\Delta K^\prime\approx5$--6~mags at $\rho\approx0$\farcs{5}. The latter corresponds to spectral types of M8.0, assuming the stellar parameters in Table~\ref{tab-prop} and using Table~5 in \citet{PM2013}, thus we can rule out any companions down through late-M types outside of 0\farcs{5}.

Our AO imaging of \hardey{} (Section~\ref{sec-ao}) revealed a second source with $\Delta K^\prime\approx6.0$~mags at $\rho\approx6$\farcs{1} with a position angle of roughly 300~degrees. Based on statistical source counts and past experience \citep{Gaidos2016} this is very likely a background star; moreover, the {\it Gaia} proper motion for this source is much smaller ($\mu_{\alpha}=-1$~mas~yr$^{-1}$, $\mu_{\delta}=-4$~mas~yr$^{-1}$) than that of \hardey{}, and the {\it Gaia} parallax puts the source at $\sim$2~kpc. There is no other indication of stellar multiplicity down to separations of $\rho\approx0$\farcs{1} with contrasts of $\Delta K^\prime\approx6$~mags, which corresponds to spectral types of M5.5, again assuming the stellar parameters in Table~\ref{tab-prop} and using Table~5 in \citet{PM2013}. Thus we can rule out companions down through mid-M types outside of 0\farcs{1}.

\subsection{Dip Properties}
\label{sec-dip}

\subsubsection{EPIC~205718330}
\label{sec-dip-330}

Figure~\ref{fig-lc1} shows the four major dimming events (A1--A4) observed during K2/C2 for \laurel{}. A1 and A3 are single dipping events, while A2 and A4 appear to be clusters of multiple dips. The dips are generally symmetric in shape given the low signal-to-noise of the data due to the faintness of the star (although one of the dips in A2 is asymmetric, it may be a blend of at least two dips). The individual dip durations are typically $\tau\simeq0.5$--1.0~days and span depths of $\delta\simeq0.5$--1.5\%. The gradients of the dips range from $\simeq$0.01--0.07~day$^{-1}$, or just a few percent of the stellar flux per day. 

\subsubsection{EPIC~235240266}
\label{sec-dip-266}

Figure~\ref{fig-lc2} shows the six major dimming events (B1--B6) observed in the K2/C11 lightcurve of \hardey{}. Similar to \laurel{}, the individual dip durations last $\tau\simeq0.5$--1.0~days. However, the depths are much shallower at $\delta\simeq0.1$\%, resulting in much smaller gradients that range from $\simeq$0.0006--0.004~day$^{-1}$. The structures of the dips in the light curve of \hardey{} are also more complicated than those of \laurel{}. In particular, some of the dips (B1, B2, B6) are accompanied by potential pre- or post-brightening events; this is likely {\em not} an artifact of the flattening process, as we can reproduce the brightening events using different normalization techniques (e.g., fitted splines, running medians). Moreover, rather than being symmetric, the dips are asymmetric---specifically, the egresses are significantly steeper (by factors of $\simeq$1.5--2.5) than the ingresses for all but the last dipping event.

%%%%%%%%%%%%%%%%%%%%%%%%%%%%%%%%%%%%%%%%%%%%%%%%%%%
%%%%%%%%%%%%%%%%%%%%% DISCUSSION %%%%%%%%%%%%%%%%%%
%%%%%%%%%%%%%%%%%%%%%%%%%%%%%%%%%%%%%%%%%%%%%%%%%%%

\section{DISCUSSION}
\label{sec-discussion}

\subsection{Ruling out Planets}
\label{sec-planet}

The dip shapes seen in Figures~\ref{fig-lc1} and \ref{fig-lc2} are clearly different from those seen for planetary transits, which have a distinct flat-bottomed shape. Additionally, the dip properties described in Section~\ref{sec-dip} can be used to rule out planet transits. For \laurel{}, the transit depths are $\delta\simeq0.5$--1.5\%, which correspond to planet radii of $R_p\simeq4$--7~$R_{\oplus}$ when simply assuming $\delta = (R_p / R_\star)^2$ and neglecting limb darkening. However, the transit durations are $\tau\simeq0.5$--1.0~days, which would require that such planets have orbital semi-major axes of $a\gtrsim1000$~AU---assuming that the transit duration goes as $\tau = (P/\pi) \arcsin( \sqrt( (R_\star + R_p)^2 + b^2) / a)$, where $P$ is the orbital period and $b$ is the impact parameter. Similar arguments can be applied to rule out planetary transits for EPIC~235240266, which had much smaller dip depths ($\sim$0.1\%) corresponding to planet radii of $R_p\sim2$~$R_{\oplus}$, but similarly long transit durations of $\tau\simeq0.5$--1.0~days.

\subsection{Gravitationally-bound planetesimal clouds?}
\label{sec-clumps}

The dimming events associated with the little dippers could instead be due to occultations of the star by clumps of dusty circumstellar material. This mechanism has been invoked to explain some of the much deeper dimming events associated with the aperiodic young dipper stars that are not significantly accreting and host evolved primordial disks \citep{Ansdell2016a}. Indeed, optically thin debris disks containing such objects would go undetected by the currently available observations of these sources (Section~\ref{sec-excess}).

To put constraints on the possible sizes and orbital distances of such objects, we followed \cite{Ansdell2016a} and \cite{Boyajian2016} to consider clumps of circumstellar material much less massive than their host star and on circular orbits:

\begin{equation}
R_{c} \approx 1.85 \, \tau \, \Big(\frac{M_{\star}}{a}\Big)^{1/2} - R_{\star},
\label{eqn-size}
\end{equation}

where $a$ is the semi-major axis of the clump orbit in AU, $R_{c}$ is the clump radius in solar radii, $\tau$ is the dip duration in days, and $M_{\star}$ and $R_{\star}$ are the stellar mass and radius, respectively, in solar units. The correlation between $R_{c}$ and $a$ from Equation~\ref{eqn-size} is illustrated in Figure~\ref{fig-clumps} for a star of $M_{\star} = 0.8~M_\odot$ and $R_{\star}=0.8~R_\odot$ with dip durations of $\tau=0.5$ and 1.0~days, comparable to \laurel{}.

\begin{figure}
\includegraphics[width=8.5cm]{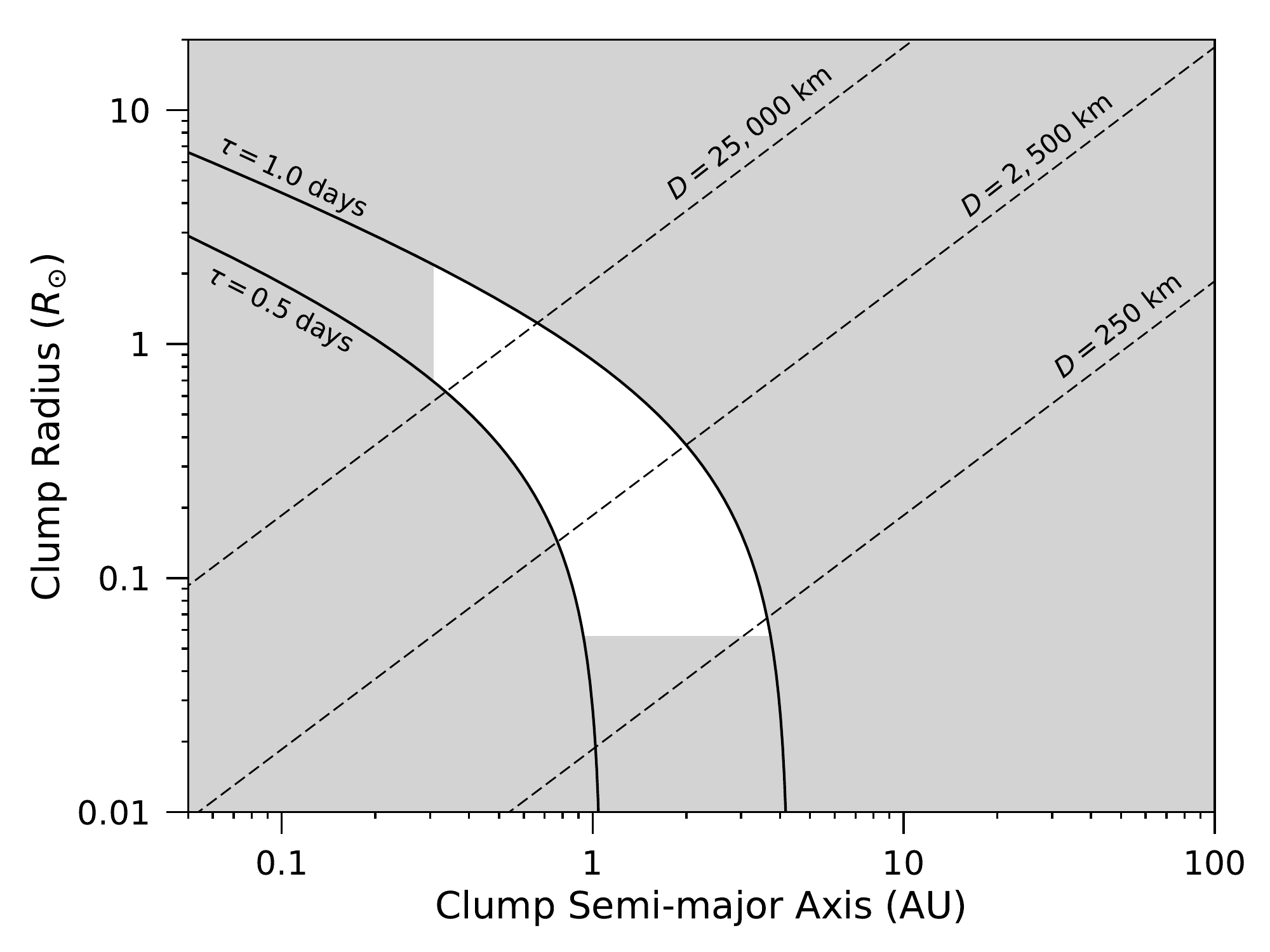}
\caption{\small Relation between clump radius ($R_{c}$) and semi-major axis ($a$) for objects orbiting \laurel{} (Section~\ref{sec-clumps}) using Eqn.~\ref{eqn-size}. Solid black lines show constant dip durations ($\tau$), assuming a host star of $M_{\star} = 0.8~M_{\odot}$ and $R_{\star} = 0.8~R_{\odot}$. Dashed black lines show Hill spheres of objects with different diameters ($D$). Gray regions show forbidden values of $a$ and $R_{c}$ based on constraints from the aperiodicity of the events, minimum dip depths of $\delta_{\rm min}\simeq0.5$\%, and dip durations of $\tau\simeq0.5$--1.0~days. 
}
\label{fig-clumps}
\end{figure}

We can place lower limits on $a$ due to the aperiodicity of the dimming events, which implies that the clumps have orbital periods longer than the $\simeq$80-day {\it K2} observing campaigns; inputting this limit into Kepler's third law gives $a>0.34$~AU. We can also place lower limits on $R_{c}$ from the minimum observed dip depth ($\delta_{\rm min}$) since the smallest corresponding $R_{c}$ would be set by an optically thick clump \cite[e.g., see Equation~5 in][]{Ansdell2016a}; using $\delta_{\rm min} = 0.5$\% (e.g., for \laurel{}) implies that $R_{c}>0.06~R_{\odot}$. These limits are shown by the gray regions in Figure~\ref{fig-clumps}. 

The clumps must be gravitationally bound in order to be stable against orbital shear \citep[e.g.,][]{KB2005} and internal velocity dispersion \citep[e.g.,][]{JW2012}. Thus we could assume they are gravitationally bound within the Hill spheres of planetesimals or planet-sized objects, where the Hill sphere is defined as $R_{\rm Hill} = \xi a \sqrt[3]{ M_{\rm p} / (3M_{\star})}$ and $\xi$ is the fraction out to which orbits are stable. Figure~\ref{fig-clumps} shows the Hill spheres of objects of various diameters, assuming bulk densities of 3~g~cm$^{-3}$ and using $\xi=0.5$ \cite[i.e., an average between that expected from retrograde and prograde orbits;][]{Rieder2016}. Given these constraints, the Hill spheres of objects with diameters $D\sim250$--25,000~km (i.e., spanning large-asteroid to super-Earth sizes) can produce the observed dips. 

One limiting factor is the required number of such objects that would need to be orbiting at a given semi-major axis for us to observe 4--6 unique transits during the $\simeq$80-day {\it K2} campaigns. As in \cite{Ansdell2016a}, we can estimate the total number of clumps that would pass through our line of sight, assuming they are evenly distributed along the orbit, and correcting for the fact that the line-of-sight only intercepts $R_\star / H$ of the disk for edge-on inclinations, where $H$ is the scale height of the disk and assuming $H/a\sim0.06$ \cite[the median for debris disks;][]{Hughes2018}. This gives the total number of clumps in the disk as:

\begin{equation}
N_{\rm tot} \approx 60\, N_{\rm obs} \, \frac{a^{5/2}}{M_{\star}^{1/2} R_{\star}},
\label{eqn-nclumps}
\end{equation}

where $a$ is in AU, and $M_{\star}$ and $R_{\star}$ are in solar units. This equation predicts a few hundred objects at $\sim$1~AU, again assuming a host star of $M_{\star} = 0.8~M_{\odot}$ and $R_{\star} = 0.8~R_{\odot}$. Given the constraints shown in Figure~\ref{fig-clumps}, these objects may have a range of diameters. Predictions for smaller orbital radii give smaller required numbers, but correspond to much larger objects. Even the smallest objects would correspond to some of the largest asteroids in our Solar System, though survival of such planetesimal disks could be stable over Gyr timescales \cite[e.g.,][]{Heng2010}. Note that this number of required objects no longer holds if the observed objects are clustered in one part of the orbit, and future monitoring of these systems will test this assumption. 

Nevertheless, the modest WISE limits on infrared excess do not clearly rule out the possibility of such a population of objects existing. For \laurel{}, $\sim$1\% of the starlight is blocked by orbiting clumps $\sim$5\% of the time (see Figure~\ref{fig-lc1}), giving an average dimming of $\sim$0.05\%. This average dimming corresponds to the expected dust fractional luminosity $f$ for a spherical cloud of clumps surrounding a star, or for a family of clumps along an orbit when corrected by the factor $\sim R_\star / a$ \citep{Wyatt2018}. The WISE limits predict $f\approx0.2$\% for \laurel{} (Section~\ref{sec-excess}), thus are clearly insufficient to rule out even a spherical cloud of clumps. For \hardey{}, the dips appear to block $\sim$0.1\% of the starlight $\sim$10\% of the time (see Figure~\ref{fig-lc2}), implying an average dimming of $\sim$0.01\%, which is still below but more comparable to the WISE limits of $f\approx0.03$\%. Thus for \hardey{} the WISE limits cannot rule out a family of clumps along an orbit and perhaps also a spherical cloud of clumps around the star.

\subsection{Star-grazing Exocomets?}
\label{sec-scatt}

Alternatively, the dipping events seen in the light curves of \laurel{} and \hardey{} could be produced by transits of comets, or fragments of a comet, undergoing very close approaches to the star. Thousands of ``Sun-grazing" and ``Sun-skirting" comets  in our own Solar System have been detected by the Solar and Heliospheric Observatory ({\it SOHO}) Large Angle Spectrometric Coronagraph \cite[LASCO;][]{Brueckner1995}, and many appear to belong to discrete comet families \citep{Jones2018}. 

\citet{Rappaport2018} identified several dips in the {\it Kepler} light curve of a main-sequence F2V-type star and explained them by occulting exocomets. These dips have distinctive asymmetric ``talon'' shapes that represent transits of trailing dust tails \citep{Lecavelier1999} as well as, in one case, a pre-transit brightening possibly due to forward scattering by dust in the trailing tail \cite[e.g.,][]{DeVore2016}. Comparable transit shapes and brightening events have also been seen in the {\it Kepler} and {\it K2} light curves of the ``disintegrating planets" \cite[see review in][]{vanLieshout2017}, which have exhibited evidence for both leading and trailing dust tails \cite[e.g.,][]{SO2015}.

Similar shapes and effects are seen in the light curves of the little dippers. Not only are the dips of similar duration ($\sim$1~day) and depth ($\sim$0.1\%), but also the egresses of \hardey{}'s dips are significantly steeper than the ingresses (Section~\ref{sec-dip-266}), suggestive of leading dust tails. Moreover, there is at least one potential pre-transit brightening event (B6 of \hardey{}), indicative of forward scattering by dust. The lack of these features in the light curve of \laurel{} does not rule out the exocomet scenario, as such distinctive shapes are not necessarily observed for some cases of orbit, star, and dust properties \citep{Lecavelier1999b}. 

Consider a single cometary body of radius $R_c$ on a near-parabolic orbit with periastron distance $q$ around a star with mass $M_\star$. At time $t$ relative to the epoch of periastron, the body will have a true anomaly $\nu$ such that:
\\
\begin{equation}
\label{eqn.time}
t = \sqrt{\frac{2q^3}{GM_\star}}\left(\tan \frac{\nu}{2} + \frac{1}{3}\tan^3 \frac{\nu}{2}\right).
\end{equation}

From Equation~\ref{eqn.time}, the rate of change of true anomaly is:

\begin{equation}
\label{eqn.trueanom}
\dot{\nu}=\sqrt{\frac{2GM_\star}{q^3}}\cos^4 \frac{\nu}{2},
\end{equation}

and thus the rate of change of distance from the star $a = 2q/(\cos \nu + 1)$ when using Equation~\ref{eqn.trueanom} is:

\begin{equation}
\label{eqn.distchange}
\dot{a} = \frac{a^2}{2q} \sqrt{\frac{2GM_\star}{q^3}}\cos^4 \frac{\nu}{2} \sin \nu.
\end{equation}

Assuming mass-loss from the comet is driven by evaporation of a dark, ice-rich surface with a specific latent heat of vaporization $C$ and containing a fraction $f_d$ of spherical dust grains of radius $s$ and density $\rho_d$, the cross-sectional surface area of dust ($A$) produced per unit time is:

\begin{equation}
\label{eqn.adot}
\dot{A} = \frac{3L_\star R_c^2f_d}{16C\rho_d s a^2}.
\end{equation}

Re-expressing Equation \ref{eqn.adot} as the surface area ejected per unit distance from the star, and using Equation~\ref{eqn.distchange}, we have:

\begin{equation}
\label{eqn.area}
\frac{dA}{dR} = \frac{3L_\star R_c^2f_dq}{16C\rho s}\sqrt{\frac{2q^3}{GM_\star}}\frac{1}{a^4 \sin(\nu) \cos^4( \nu/2)}.
\end{equation}

The $a$ and $\nu$ dependencies in Equation~\ref{eqn.area} show that dust production will strongly peak at periastron. Since the probability of a transiting geometry scales as $1/a$, these objects are also most likely to significantly obscure the host star near periastron. Moreover, since the flux contributed to the emission in some infrared band-pass will also increase with temperature around periastron, essentially all of the contribution to excess infrared flux is from dust ejected around periastron, assuming the body survives the passage.  

We now assume that, during the transit (near periastron), the stellar wind imparts a component to the dust velocity \emph{towards} the observer, and that the transverse velocity remains approximately the Keplerian value. Thus the total cross-sectional area of dust is the production rate (Equation~\ref{eqn.adot}) multiplied by the transit duration $\tau = 2R_\star/v(q)$, where $v(q)$ is the transit velocity at periastron. The maximum dip depth (if the dispersion normal to the orbital plane is small compared to the stellar radius) is therefore: 

\begin{equation}
\label{eqn.depth}
\delta^* = \frac{3L_\star R_c^2f_d}{16 \pi C\rho s q R_\star}\sqrt{\frac{2}{GM_\star q}}.
\end{equation}

For a 50-50 water ice-dust composition, $f_d=0.5$ and $C$ is half that of water ice, thus we estimate from Equation~\ref{eqn.depth}:

\begin{equation}
\label{eqn.depth2}
\delta^* \approx 8 \times 10^{-8} \left(\frac{R_c}{1 {\rm km}}\right)^2\left(\frac{s}{1 \mu m}\right)^{-1}\left(\frac{q}{1 {\rm au}}\right)^{-3/2}\frac{L_\star}{L_{\odot}}\left(\frac{M_\star}{M_{\odot}}\right)^{-1/2}\left(\frac{R_\star}{R_{\odot}}\right)^{-1}.
\end{equation}

Hence for a periastron distance of $q = 0.05$~AU \cite[Sun-grazing Kreutz-family comets exhibit a maximum in brightness at this distance;][]{Knight2010} around a main-sequence K dwarf star with $M_\star = 0.8M_{\odot}$, $R_\star = 0.8R_\odot$, and $L_\star = 0.4L_{\odot}$ (i.e., similar to \laurel{}), the required comet radius to produce $\sim$1\% dips with $\sim$1~$\mu$m grains is about 50~km.  While perhaps common in very young planetary systems, objects of such size would not be expected to be common among older stars. On the other hand, tidally or thermally induced disruption of a 10~km size body into $\sim$100~m fragments, analogous to that occurring for Sun-grazing and Sun-skirting comets, would produce the required surface area. The potential clustering of the observed dipping events supports this scenario, although this disruption must happen very near periastron to explain the discrete appearance of the dips.  The total mass of dust involved to produce the dips for the above case is in fact what would be expected for a 10-km radius comet.   

Likewise, we can approximate the infrared flux as that coming from dust expelled over the time of periastron approach as emission $B_{\lambda}$ (e.g., blackbody) with corresponding equilibrium temperature $T(q)$ at periastron. Since $a < 2q$ within $-\pi/2 < \nu < \pi/2$, we can approximate this close-approach time $\tau_q$ using Equation~\ref{eqn.time} as:

\begin{equation}
\label{eqn.emit_time}
\tau_{q} \approx \frac{8}{3}\sqrt{\frac{2q^3}{GM_\star}}.
\end{equation}

Combining Equation~\ref{eqn.adot}, \ref{eqn.depth}, and \ref{eqn.emit_time} we arrive at:

\begin{equation}
F_{\lambda} = \frac{8\pi}{3}B_{\lambda}(T(q))R_\star q \delta^*.
\end{equation}

The infrared excess as a fraction of the stellar photosphere emission is then:

\begin{equation}
f_{\lambda} = \frac{8qB_{\lambda}(T(q))}{3R_\star B_{\lambda}(T_\star)}\delta^*.
\end{equation}

Even for very hot ($\sim$1000~K), large ($s\gg1$~$\mu$m) dust grains that emit efficiently, the expected 4.6~$\mu$m excess is $f \sim 2\delta^* \lesssim 2$\%. This is similar in magnitude to the errors in the WISE photometry (Table \ref{tab-phot}), and the infrared signature of a \emph{single} such object would have gone undetected.  Moreover, the epoch of WISE observations is not the same as \emph{K2} and the time scale for the dust to be produced and blown away from the star is very short compared to the elapsed time between the two sets of observations.  

Additionally, the differences in the depths and durations of the dipping events for \laurel{} and \hardey{} are consistent with objects disintegrating at a common stellar irradiance. The depths of the dips for \laurel{} are $\sim$1\% while those for \hardey{} are $\sim$0.1\%; the ratio of these dip depths is consistent with the square of the ratio of their stellar radii, which is at least $\simeq$5. Moreover, the dips for \hardey{} last roughly 1 day while those for \laurel{} are roughly 0.5 days, consistent with the approximation that the difference in duration for a comet disintegrating at a certain irradiance will scale as $\tau \propto \sqrt{a/M_\star}$ where $a\propto \sqrt{L_\star}$ thus $\tau \propto L_\star^{1/4} M_\star^{-1/2}$.  In other words, the dipping events seen in \laurel{} and \hardey{} are consistent with similar objects disintegrating at an irradiance corresponding to distances at which Sun-grazing comets are observed to break up.  At least some Sun-grazing and Sun-skirting comets are believed to have evolved onto their close-approaching orbits via a Kozai resonance interaction with the Sun and Jupiter \citep{Jones2018}. Thus it is reasonable to speculate that, if these little dippers are indeed star-grazing exocomets, they indicate the existence of a massive exoplanet that does not transit but might be revealed by long-term radial velocity monitoring.

%%%%%%%%%%%%%%%%%%%%%%%%%%%%%%%%%%%%%%%%%%%%%%%%%%%
%%%%%%%%%%%%%%%%%%%%% CONCLUSION %%%%%%%%%%%%%%%%%%
%%%%%%%%%%%%%%%%%%%%%%%%%%%%%%%%%%%%%%%%%%%%%%%%%%%

\section{CONCLUSION}
\label{sec-conclusion}

We presented two systems, \laurel{} and \hardey{}, which based on a visual survey of the {\it K2} dataset appear to represent a class of objects that are rare---either intrinsically or due to the limited sensitivity and time baseline of {\em K2} observations. The light curves of these stars contain episodic drops in flux with profile shapes and durations similar to those of the young ``dipper" stars (i.e., a mixture of symmetric and asymmetric dips lasting $\simeq$0.5--1.0 days), yet with depths $\sim$1--2 orders of magnitude shallower (i.e., $\simeq$0.1--1.0\% in flux). We conducted ground-based follow-up of these ``little dippers" to derive their stellar parameters and place constraints on the possible mechanisms causing the dimming events.

We first vetted the {\it K2} data of \laurel{} and \hardey{} for instrumental and/or data processing artifacts that could be causing the dimming events. We tested different parameters for K2SFF de-trending, ruled out contamination from ``rolling bands" and nearby bright stars, and checked for evidence of CCD ``cross-talk." Based on these tests, we concluded that the dipping events were both astrophysical and associated with the target sources. 

Our follow-up observations and analysis showed that these stars are clearly not young. Their lack of detectable infrared emission in excess of the stellar photosphere precludes protoplanetary disks, though cannot rule out debris disks due to the limited WISE photometric precision. The lack of Li {\sc I} absorption also puts limits on their age of $\gtrsim150$~Myr and $\gtrsim800$~Myr for \laurel{} and \hardey{}, respectively. Finally, their galactic space motions are inconsistent with the typical $UVW$ values for young disk stars, signaling that they are kinematically old. Follow-up spectra indicate that \laurel{} and \hardey{} are likely early-K and late-F type dwarfs, respectively, though \hardey{} could be a K-type subgiant.

Given these constraints, we explored two mechanisms for explaining the dimming events: dust-enshrouded remnants of planet formation and star-grazing exocomets. We argued that these little dippers are consistent with transits of star-grazing exocomets due to: (1) the shape and depth of the dipping events being consistent with a disrupted comet-sized object transiting while at a periastron distance very similar to those of Sun-grazing comets seen in our own Solar System, and (2) the differences in the depths and durations of the dipping events for \laurel{} and \hardey{} being consistent with similar objects disintegrating at a common stellar irradiance. We speculated that this could indicate the existence of massive non-transiting exoplanets driving the close-approach orbits. 

One question is why we do not see more of these exocomet-related events in the {\it Kepler} and {\it K2} datasets. \cite{Rappaport2018} suggested that the rarity of the exocomets in the {\it Kepler} dataset could be due to our current photometric precision being only sufficient to detect the largest (and rarest) exocomets with the transit method; smaller exocomet systems, perhaps more typical around the older {\it Kepler} stars, could still exist but go undetected. However, one might then expect that larger exocomet bodies would be more common in the younger ($\sim$10--100~Myr) systems surveyed by {\it K2}. It is therefore possible that such large exocomets are inherently rare, even in young systems. Another possibility is that, if these really are star-grazing exocomets driven by massive perturbing outer planets, then the low occurrence of gas giants seen in the exoplanet population, combined with the low probability of observing transits of objects on highly eccentric orbits, could explain the rarity of the little dippers. Additional exocomet-like systems could be found by the recently launched Transiting Exoplanet Survey Satellite \cite[TESS; ][]{Ricker2015}; building up a larger sample and identifying common properties or trends will be important for understanding these objects.

%%%%%%%%%%%%%% ACKNOWLEDGEMENTS %%%%%%%%%%%%%%%%%%

\section*{Acknowledgements}

MA acknowledges support from the Center for Integrative Planetary Science at the University of California at Berkeley as well as grants NSF AST-1518332, NASA NNX15AC89G and NNX15AD95G/NEXSS. CFM is supported by an ESO Fellowship. GMK is supported by the Royal Society as a Royal Society University Research Fellow. AV's work was performed under contract with the California Institute of Technology/Jet Propulsion Laboratory funded by NASA through the Sagan Fellowship Program executed by the NASA Exoplanet Science Institute. This work is based on observations collected at the European Organisation for Astronomical Research in the Southern Hemisphere under ESO programme 0101.C-0866. This work has made use of data from the European Space Agency (ESA) mission {\it Gaia} (\url{https://www.cosmos.esa.int/gaia}), processed by the {\it Gaia} Data Processing and Analysis Consortium (DPAC, \url{https://www.cosmos.esa.int/web/gaia/dpac/consortium}). Funding for the DPAC has been provided by national institutions, in particular the institutions participating in the Gaia Multilateral Agreement. We made use of the following Python packages: Astropy \citep{Robitaille13}, a community-developed core Python package for Astronomy, and Matplotlib  \citep{Hunter07}, a plotting package used to construct the figures in this paper. This work benefited from NASA's Nexus for Exoplanet System Science (NExSS) research coordination network sponsored by NASA's Science Mission Directorate. We are grateful to Tae-Soo Pyo and Akihiko Fukui for their support with IRCS observations and data reduction. This work was supported by JSPS KAKENHI Grant Number JP16K17660. TJ and DL acknowledge developer Allan R. Schmitt for making his lightcurve examining software {\tt LcTools} freely available.

%%%%%%%%%%%%%%%%%%%% REFERENCES %%%%%%%%%%%%%%%%%%

\bibliographystyle{mnras}
%\bibliography{ms} 

%%%%%%%%%%%%%%%%%%%% END %%%%%%%%%%%%%%%%%%%%%%%%%

% Don't change these lines
\bsp	% typesetting comment
\label{lastpage}
\end{document}